\documentclass[twocolumn,secnumarabic,endfloats,aps]{revtex4}

\usepackage{graphicx}
\usepackage{dcolumn}
\usepackage{bm}

\begin{document}

\preprint{APS/123-QED}

\title{Correlating elasticity and crack formation}

\author{$^1$Petr Lazar}
\author{$^1$Raimund Podloucky}
\author{$^2$Walter Wolf}

\affiliation{%
 $^1$ Institut f\"ur Physikalische Chemie, Universit\"at Wien,
 Liechtensteinstrasse 22A, A-1090, Vienna, Austria\\
 $^2$ Materials Design s.a.r.l., 44, av. F. -A. Bartholdi, 72000 Le Mans, France}%
\date{\today}

\begin{abstract}
For solving the longstanding materials science problem of correlating elastic
properties of a solid material to the formation of cracks we present a new
general concept.  This concept is applied to the technologically most
important cracks of loading mode I for which we establish exact correlations
by introducing  a localization length as a new material parameter. We study
two limiting cases of crack formation making use of analytic models
determining the material and direction dependent parameters by comparison to
{\em ab initio} density functional results.  This is done for a variety of
real materials in order to test our approach for different types of bonding.
For the most interesting ideal brittle cleavage we find that
the localization length is -within a reasonable approximation- of constant
value, which results in a simple relation for the critical stress, presumably
being useful for materials engineering.  Our results confirm that the proposed
general concept results in meaningful physical models which -for the first
time- allow a rigorous and simple correlation between elastic and mechanical
properties.
\end{abstract}

\maketitle

\section{\label{sec:level1} Introduction}
It is a long standing effort in materials science to describe or just to
estimate the critical property of crack formation of a solid material in terms
of its elastic properties, an objective both of scientific as well as
technological interest. Although more than 80 years have been passed since the
pioneering work of Griffith~\cite{Griffith}, it seems that still there is a
need for quantitative models on the atomic scale in order to establish a correlation between
elastic and mechanical properties.  First attempts for cracks of mode I
estimating the critical cleavage stress were made by  Polanyi~\cite{Polanyi}
and Orowan~\cite{Orowan}.  Further studies~\cite{Gilman,Kelly} 
tried to improve Orowan's approach but the quantitative
results were not satisfying.  Very recently, Hayes {\em et al.}~\cite{Hayes}
-based on the  work of Nguyen and Ortiz~\cite{Nguyen}-  applied a
rather complicated concept,  which for intrinsic properties on the atomic scale 
involves a dependency on the macroscopic thickness of the material.

The conceptual problem of correlating elastic and cleavage properties  consists
in correlating  a {\em non-local} property to a {\em local} property: the
elastic response to a perturbation is {\em non-local} because the energy of
is distributed over a macroscopic volume $V_{mac}$ of the
whole material, whereas the energy for the formation of a crack  is considered
to be {\em localized} in some local volume $V_{loc}$ in the vicinity of the
crack.  Let us now imagine the initialization of an infinitesimally small crack 
into the solid at equilibrium.  This
perturbation may either open a finite crack or the solid may respond in a purely
elastic way by distributing the energy uniformly over its macroscopic
dimensions, which constitutes some kind of unstable equilibrium. Then,
one can set equal the elastic response and the crack formation for
infinitesimally small crack sizes.
Consequently, a correlation  between the energies (both, elastic as well as
cleavage energies) in the {\em localized} volume $V_{loc}$, or -optionally- in
the {\em non-local} macroscopic volume $V_{mac}$ should exist.  This concept
of an unstable equilibrium should be suitable for general modes of crack
formation and elastic responses. It might, however, be very difficult to
derive a correlation in analytic form.

\begin{figure}
\includegraphics[scale=0.6]{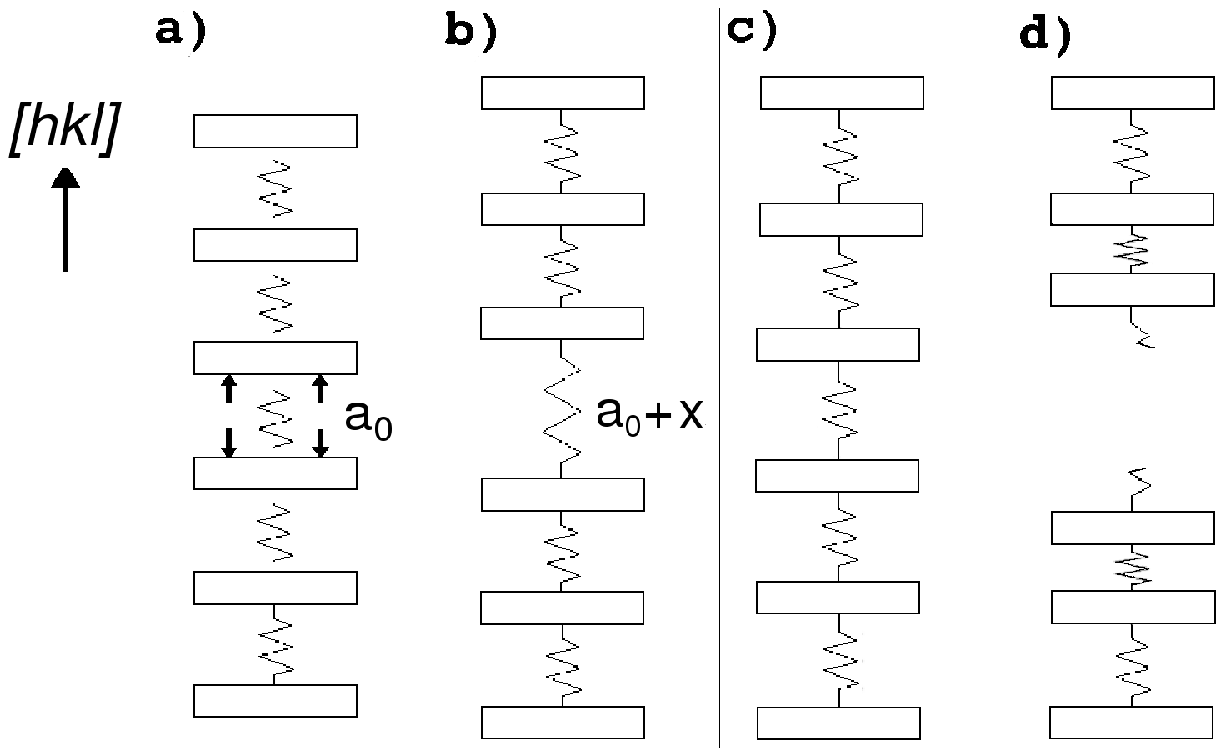}
\caption{\label{fig1} Models for cracks of mode I. A solid (sketched as a
stacking of interacting layers with a layer distance $a_0$, panel a) undergoes
brittle cleavage as sketched in panel b: a crack of size $x$ cleaves the
material into two rigid blocks without relaxing the atomic geometry of the
layers in the blocks.  For the ideal elastic cleavage (see text) the
material relaxes in a perfectly elastic manner (panel c) up to a critical
crack size above which it breaks abruptly into two blocks with atomically
relaxed surfaces (panel d).}
\end{figure}

At least for simple crack modes an analytic formulations exists.
We demonstrate this for cracks of  mode I as sketched in
Fig. \ref{fig1}: along a plane with fixed area $A$ and orientation $[hkl]$ 
a single crystal is cleaved into two blocks.  The local volume
$V_{loc}$ can then be expressed by 
\begin{equation} \label{eq1} 
V_{loc} = A \,\, L,
\end{equation}
with $L$ being a finite length in which the energy is localized.
Consequently, the macroscopic volume  $V_{mac}$ is defined as
\begin{equation}
\label{eq2} V_{mac} = A \,\, D,
\end{equation} 
where $D$ describes the macroscopic thickness of the material. By Eqs.
\ref{eq1} and \ref{eq2} a simple rescaling condition between the two volumes
exists, namely $V_{mac} = V_{loc} (D/L)$. The rescaling factor $D/L$ (or
its inverse) transforms local quantities into non-local ones (or vice versa
for the inverse factor). We will make use of the conversion factor
when we consider two extreme models of crack formation: a)
the ideal brittle case modelling cleavage of the material into two
rigid blocks, i.e. the atomic positions within the blocks do not relax. This
process corresponds to a  crack formation which is so {\em fast} that the atomic
positions do not relax during the process (see panel b of Fig. \ref{fig1}); 
b) a process we will
call ideal elastic crack formation,  which  is so {\em
slow} that
atoms will always fully relax after a crack with a given
size $x$ has been openend.  If $x$ is larger than a critical
value then the material should abruptly break into two blocks with geometrically relaxed
surfaces (as sketched in panels c and d of Fig. \ref{fig1}).
For both cases, analytical models for the crack formation energies can be
formulated. The yet unknown parameters of these models will be determined by
fitting the analytic expressions to energies obtained from density functional
theory
(DFT) calculations modelling the  processes a) and b). For this purpose we
make use of the Vienna Ab initio Simulation Package  \cite{VASP2} for
calculating the total DFT energies and related quantities for suitably large
slabs. Atomic relaxations are determined by minimizing the forces acting on
the atoms.  

The final goal, the correlation between elastic properties and
crack formation, is then formulated for infinitesimally small crack sizes and
elastic strains by setting equal crack formation energy and elastic energy,
and choosing either a local or non-local description.  In order to demonstrate
the generality of our concept we apply it to a variety of materials
representing different types of bonding, such as metallic, covalent and ionic
bonding.

\section{\label{brittle} Ideal Brittle cleavage}

As an analytical model for ideal brittle cleavage we choose the so-called
universal binding relation (UBER) \cite{uber} for the  energy
$E_b(x)$ depending on the crack size $x$,
\begin{equation}\label{eq3}
E_{b}(x) = G_b \left[ \left( 1 + \frac{x}{l_b} \right) \exp \left( -
\frac{x}{l_b} \right) - 1 \right] ,
\end{equation}
which represents some kind of interatomic potential.  
The material and
direction dependent parameters are the cleavage energy $G_b$ and the critical
length $l_b$ at which  the stress $\sigma(x)=dE/dx$ reaches a maximum (the index $b$
denotes brittle cleavage). The UBER energy behaves
quadratically at $x=0$, and for very large separations $x$ it approaches the
energy zero with a negative curvature.  It should, however, be mentioned that
the UBER is not universal as usually claimed 
(see e.g. Refs.~\cite{uber,Hayes}) because the curvature for
$E_b({x\rightarrow \infty})$ could also be positive~\cite{note1}. Furthermore,
the two material parameters $G_b$ and $l_b$ might also depend on the actual
planes between which the crack is initialized if there are different
layer-to-layer distances  along the direction $[hkl]$ (e.g. for the diamond
lattice in $[111]$ direction).

In the case of ideal brittle fracture the atomic positions remain
fixed and do not relax. The electronic states, however, will change due to
the change of the atomic environment in the neighbourhood of the crack because
potentials at the surface will
be different compared to the bulk environment. Directional bonds might be reordered
which is reflected in the change of the electron distribution and,
consequently, the energy
of the system. All the electronic relaxation effects are fully taken into
account by
the self-consistency of the DFT approach which is used to fit the UBER.  
For this purpose we
calculate DFT total energies $E_{DFT}(x_i)$ for a suitably large number of
different crack sizes $x_i$ and for sufficiently large slabs.

Studying UBER for a variety of different materials with different types of
bonding, we found that for most cases the fit  to the DFT values is rather
good and therefore the fitted parameters $G_b$ and $l_b$ listed in Table
\ref{table1} are physically meaningful.  
Deviations are found for crack sizes
$x > l_b$ which could be particularly large when strong covalent bonds are
broken.  Anyway, the shortcomings of UBER are not of significant importance
for the present study.

If the response to the perturbation is elastic, the energy of the perturbation
is
expressed by
\begin{equation}\label{eq4}
E_{elast}(x) = \frac{1}{2} C V \delta^2,
\end{equation}
which comprises the uniaxial modulus $C$ for straining the material along
$[hkl]$ with a fixed area $A$ of the planes perpendicular to $[hkl]$.
The symbol $V$ denotes the volume
over which the energy is uniformely distributed,
and  $\delta$ is the dimensionless relative strain. 
If we decide to refer to a {\em local}
description then $V$ is  substituted by V$_{loc}$, 
and the relative strain is  defined by $\delta_{loc} = x / L_b$ 
with $x$ being the initial very small crack size.
The localization length for brittle cleavage $L_b$ which is an intrinsic parameter,
depends on the material, direction and cleavage planes like $G_b$ and $l_b$.
For very small  $x$ we set now equal the elastic energy
$E_{elast}$ and the UBER expression (i.e. the Taylor expansion of $E_b(x)$ in
Eq. \ref{eq3} in lowest order of $x$~\cite{Vogtenhuber}), which results in the relation
\begin{equation}\label{eq5}
\frac{1}{2} G_b \frac{x^2}{l_b^2} = \frac{1}{2} A L_b C \frac{x^2}{L_b^2} .
\end{equation}
This is the key equation to correlate the ideal brittle cleavage and its material
parameters $G_b, l_b,$ and $L_b$ to the elastic properties
described by the uniaxial modulus $C$. 
From Eq.~\ref{eq5} a variety of
relations can be derived. 
For the critical stress, i.e. the maximum $\sigma_b:=\sigma(x=l_b)$, one obtains the  equation 
\begin{equation}\label{eq6}
\frac{\sigma_b}{A} = \frac{1}{e}\frac{G_b}{A}\frac{1}{l_b}\, = \, 
\frac{1}{e}\frac{l_b}{L_b}C \, .
\end{equation}

\begin{figure} 
\includegraphics[scale=0.7,clip]{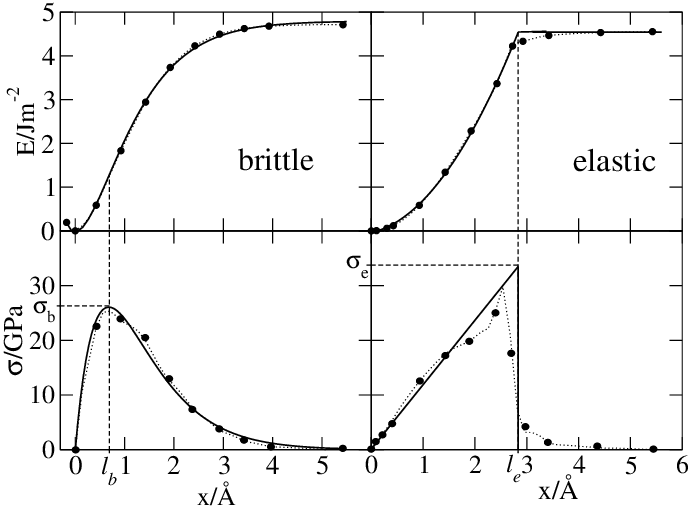} 
\caption{\label{fig2} 
Cleavage along $[100]$ of NiAl:
energy  and stress per surface area  vs. crack size $x$ for
the ideal brittle (left panel) and ideal elastic (right panel) cleavage model
full lines: analytic models, dots: DFT results).} 
\end{figure}

\section{\label{elastic} Ideal elastic cleavage}

In contrast to the ideal brittle case we now consider a cleavage process for
which
the crack formation is so slow that -after opening a crack of size $x$- the
atoms have sufficient time to fully relax. Furthermore, we require that the
material should behave perfectly elastic up to a
critical crack size $x=l_e$, at which it breaks abruptly without any
further interaction between the cleaved blocks, as sketched in Fig. \ref{fig1}
(panels c and d). We will call this process 
ideal elastic cleavage and we characterize the corresponding parameters by the index
$e$. Denoting the corresponding cleavage energy by $G_e$ we derive the obvious equation
\begin{equation} \label{eq7}
E_e(x) = \frac{G_e}{l_e^2} x^2
\end{equation}
for the corresponding cleavage energy $E_e$ for crack sizes  $x \leq l_e$. For
crack sizes $x > l_e$ we require the condition $E_e(x)=G_e$.  Of course, this
is  an idealized model because a real material will deviate from the simple
ideal elastic behaviour at least close to the critical crack size, as can be
seen in the right panel of Fig. \ref{fig2} in which the analytic model is compared to a
"real" material as calculated by the DFT procedure.

Again, like for the brittle case we establish the correlation between elastic and
cleavage properties by setting equal elastic and cleavage energy
for very small crack sizes $x$. In order to demonstrate the rescaling feature
of our concept (as discussed at the beginning)
we want a formulation in terms
of a {\em non-local} description: the strains are now defined with respect to
the macroscopic thickness by $\delta=x/D$ as in standard elastic theory.
Doing that, we have to rescale  the cleavage
energy by the obvious factor $L_e/D$. Now we have again introduced
a localisation length as a new material parameter, and
we can derive the key equation for correlating elastic and cleavage properties
for the ideal elastic case,
\begin{equation}\label{eq8}
G_e \frac{L_e}{D} \frac{x^2}{l_e^2} =  \frac{1}{2} A D C \frac{x^2}{D^2}.
\end{equation}
It should be noted that in this equation the macroscopic thickness $D$ cancels
out and, because of this cancellation, the relation contains
only  intrinsic material parameters.
In fact, we would have obtained the same relation if  the strain
would be defined relative to the localization length by
$\delta=x/L_e$, similar to treatment of the ideal brittle cleavage. 

Exploiting Eq.~\ref{eq8} we derive relations involving the
critical stress $\sigma_e:=\sigma(x=l_e)$, 
\begin{equation}\label{eq9}
\frac{\sigma_e}{A} = 2\frac{G_e}{A}\frac{1}{l_e} \, = \, 
\frac{l_e}{L_e}C .
\end{equation}

The very close formal relationship to Eq. \ref{eq6} can clearly be seen.
However, it should be noted that the material dependent cleavage  quantities such as $G,
l$ and $L$ are different for both cases (as denoted by the indices $b$ and $e$), and the
constant prefactors are also different.

\section{\label{results} Results and Discussion}
In Fig. \ref{fig2} for the $[100]$ cleavage of NiAl
the energies and stresses derived from the analytic models are
compared to  DFT results. In general for the brittle case (left panels), 
the agreement is good for smaller $x$ whereas for $x > l_b$  small but
characteristic deviations occur (note also the remarks in Ref~\cite{note1}).

The ideal elastic cleavage energy (right top panel) also fits quite well to the DFT data.
Only close to the critical crack size $x \approx l_e$ deviations from the simple
analytic model are significant because then atomic relaxations at the crack surfaces
are becoming important. The differences are naturally more pronounced for the
stress, because it is the first derivative of the cleavage
energy.

\begin{figure}
\includegraphics[clip]{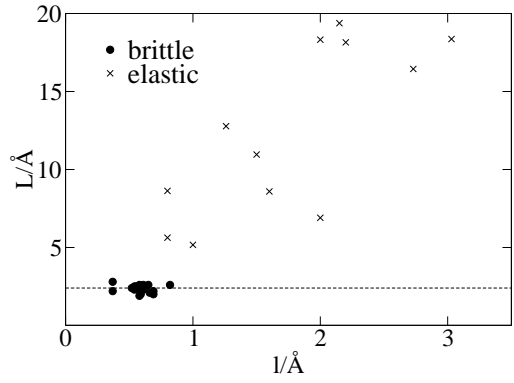}
\caption{\label{fig3} Localization lengths vs. critical lengths for ideal
brittle and elastic cleavage for a variety of materials and directions.}
\end{figure}
Studying the results for the new material parameter,
the localisation length $L$, we find a rather striking
property for the brittle cleavage as shown in Fig. \ref{fig3} and in Table
\ref{table1}, namely that -within an error of $\approx 20\%$- 
the localization lengths, $L_b \approx 2.4$ \AA\ , are rather constant for all studied materials
independent of the directions $[hkl]$ and the layer separations $a_0$.
Based on this finding we derive from
Eq. \ref{eq6} the simple relation
\begin{equation}
4.21 \, \frac{\sigma_b}{A}\sqrt{\frac{A}{G_b}}  \approx  \sqrt{C} ,
\end{equation}
in which cleavage properties (left side) are separated from
elastic properties (right side).  This equation might serve as a starting
point for the development of an  "engineering"
tool to estimate the critical cleavage properties in terms of the
uniaxial modulus C which is a linear combination of the elastic constants
\cite{note2,Nye}. 
\begin{table}
\caption{\label{table1} Parameters for brittle cleavage for some selected
compounds and directions:
uniaxial elastic modulus C (GPa), cleavage energy per
surface area $G_b/A$ (J/m$^2$), critical length $l_b$ (\AA\, ), maximum stress
$\sigma_b/A$ (GPa), interlayer distance between cleaved planes $a_0$ (\AA\,), 
and localization length $L_b$ (\AA\, ) for selected
compounds and cleavage directions $[hkl]$. In brackets, the crystal structure is denoted in
terms of Pearson's notation.}
\begin{ruledtabular}
\begin{tabular}{ccccccccc}
 & &[$hkl$]  & $C$ & $G_b/A$  & $l_b$ & $\sigma_{b}/A$ & $a_0$ & $L_b$ \\
          \hline
NiAl&(B2) & 100 & 203 & 4.8 & 0.69 & 26 & 1.45 & 2.0 \\
     &    & 110 & 284 & 3.2 & 0.54 & 22 & 2.05 & 2.5  \\
     &    & 111 & 311 & 4.1 & 0.58 & 26 & 0.84 & 2.4 \\
Ni$_3$Al&(L1$_2$) 
          & 100 & 225 & 4.3 & 0.66 & 24 & 1.78 & 2.1 \\
     &    & 111 & 331 & 3.7 & 0.52 & 26 & 2.06 & 2.4 \\
VC  &(B1) & 100 & 647 & 3.2 & 0.37 & 32 & 2.08 & 2.8 \\
     &    & 110 & 585 & 7.0 & 0.55 & 46 & 1.47 & 2.5 \\
     &    & 111 & 564 & 9.9 & 0.58 & 63 & 1.20 & 1.9 \\
MgO&(B1)  & 100 & 299 & 1.8 & 0.37 & 18 & 2.11 & 2.2 \\
     &    & 110 & 345 & 4.4 & 0.54 & 30 & 1.22 & 2.3 \\
\end{tabular}
\end{ruledtabular}
\end{table}

\begin{table}
\caption{\label{table2} Ideal elastic cleavage: Similar quantities as in Table
\ref{table1}.}
\begin{ruledtabular}
\begin{tabular}{ccccccc}
    & [$hkl$] & $C$& $G_e/A$ & $l_e$ &  $\sigma_e/A$ &$L_e$ \\
     \\ \hline
NiAl     & 100 & 203 & 4.6 & 2.7 &  34 & 15.8 \\
         & 110 & 284 &3.1 & 2.0 &  32 & 17.7 \\
         & 111 & 311 &3.9 & 2.2 &  36 & 18.4 \\
Ni$_3$Al & 100 & 225 &3.0 & 2.2 &  27 & 18.2 \\
         & 111 & 331 &3.2 & 1.0 &  64 & 5.2  \\
VC       & 100 & 647 &2.4 & 0.8 &  60 & 6.5  \\
         & 110 & 585 &6.0 & 1.6 &  75 & 12.5 \\
         & 111 & 564 &8.4 & 1.6 &  105& 8.6   \\
MgO      & 100 & 299 &1.7 & 0.8 &  42 & 5.3  \\
         & 111 & 345 &10.2 &1.9 &  107& 6.9   \\
\end{tabular}
\end{ruledtabular}
\end{table}
For the ideal elastic cleavage the critical lengths $l_e$ and
localization lengths $L_e$ are much larger than for the brittle case, as shown
in Table \ref{table2} and Fig.~\ref{fig3}. This seems to be obvious
because for the ideal elastic cleavage the material is now allowed to relax after the crack initialization
and therefore  it needs much
larger crack sizes to break it. We notice also
a strong variation of $L_e$  which is in contrast to the
brittle case. Some -but no simple-  correlation
between $l_e$  and $L_e$ exists because, generally, for larger $l_e$ the values of $L_e$ are also larger. 
More studies are needed in order to -hopefully-
find useful correlations between the material parameters for the ideal
elastic cleavage.
Also the critical strengths $\sigma_e$ are significantly enhanced from 20\% to
100\% in comparions to $\sigma_b$, whereas the cleavage energies $G_e$
-although reduced compared to $G_b$- differ not very strongly from the ideal brittle case (with
the exception of VC $[100]$).

Summarizing, we presented a general concept, which
correlates elastic and cleavage properties of a solid.
By its application to two different models for cracks of mode I
analytic and quantitative correlations were derived by
introducing a localization length $L$ , a new intrinsic material parameter which proved to be
meaningful for a large variety of materials. 
Future efforts should focus on the fundamental understanding of $L$.
\acknowledgments

{Work supported by the Austrian Science Fund in terms of the
Science College Computational Materials Science, project nr. WK04.
Calculations were performed on the Schr{\"o}dinger-2 PC cluster of the
University of Vienna.}

\begin{thebibliography}{99}
\bibitem{note1}
This occurs e.g. when cleaving diamond along $[111]$ between planes with the
shortest layer-to-layer distance because of re-ordering of strong covalent
bonds.
\bibitem{note2}
e.g. for a cubic structure the uniaxial modulus for the direction $[hkl]$ is
given by: $C=c_{11}-2(c_{11}-c_{12}-2c_{44})(h^2k^2+h^2l^2+k^2l^2)$
\end{thebibliography}

\printtables*

\printfigures*

\end{document}